\documentstyle[sprocl]{article}

\def\be{\begin{equation}}
\def\ee{\end{equation}}
\def\bea{\begin{eqnarray}}
\def\eea{\end{eqnarray}}

\begin{document}

\title{Magnetic Moment Density from Lack of Smoothness of the Ernst Potential}

\author{L. Fern\'andez-Jambrina}

\address{Departamento de Geometr\'{\i}a y Topolog\'{\i}a,\\ Facultad de
Ciencias Matem\'aticas,\\ Universidad Complutense de Madrid\\ E-28040-Madrid,
Spain}


\maketitle\abstracts{ In this talk it is shown a way for constructing magnetic
surface sources for  stationary axisymmetric electrovac spacetimes possessing a
non-smooth electromagnetic Ernst potential. The magnetic moment density is
related to this lack of smoothness and its calculation involves solving a
linear elliptic differential equation. As an application the results are used
for constructing a magnetic source for the Kerr-Newman field.}
  
\section{Introduction}

The Darmois' junction conditions \cite{Darmois} provide a way for constructing
surface sources for non-smooth spacetimes. Whenever the extrinsic curvature of
a hypersurface contained in the spacetime is discontinuous, a  distributional
stress tensor can be assigned to it \cite{ISR}. Furthermore, if the
electromagnetic curvature is discontinuous across a hypersurface, a
distributional electromagnetic source can be located on it \cite{is1}. The
existence of a timelike (axial) Killing field allows the calculation of the
mass (angular momentum) surface density in terms of the discontinuity of the
extrinsic curvature of the hypersurface \cite{is2}. An expression for the
electric charge density is also obtained for stationary fields \cite{is1}, but
so far, no expression has been provided for the magnetic moment density. In the
next section we shall follow a different approach. Instead of considering the
discontinuities of the electromagnetic curvature we shall study the
discontinuities of the electromagnetic Ernst potential \cite{Ernst},
restricting ourselves to stationary axisymmetric electrovac spacetimes. This
approach follows closely the classical theory of potential, where a
discontinuous (non-smooth) scalar potential gives rise to a dipole (monopole)
source for the field that derives from the potential. As an example, we shall
apply the formalism to the Kerr-Newman spacetime.

\section{Magnetic moment surface density} As it has already been stated, we
shall just study stationary axisymmetric electrovac spacetimes. Let us consider
the metric, 

\begin{equation}ds^{2}=-e^{2U}(dt-A\,d\phi)^{2}+e^{-2U}\{e^{2k}(d\rho^{2}+dz^{2})+\rho^2
d\phi^{2}\}\label{eq:can}, \end{equation} written in Weyl coordinates. $U$,
$A$, $k$ are functions of $\rho$ and $z$.

The electric, $E$, and magnetic field, $B$, as viewed from an orthonormal
coframe \cite{second} whose timelike one-form is given by
$\theta^0=e^U\,(dt-Ad\phi)$, can be written, according to Maxwell equations, in
terms of a complex function, $\Phi$, 

\begin{equation} f=E+i\,B=-e^{-U}\,d\Phi\label{eq:f}, \end{equation} the Ernst
electromagnetic potential \cite{Ernst}. This scalar potential $\Phi$ satisfies
one of the Ernst equations \cite{Ernst}, which in this notation can be written
as,

\begin{equation}
L\,\Phi\equiv\frac{1}{\sqrt{g}}\,\partial_\mu\,\left\{N\,\sqrt{g}\,\left(e^{-2\,U}\,
g^{\mu\nu}-\frac{i}{\rho}\,A\,\epsilon^{\mu\nu}\right)\,\partial_\nu\,
\Phi\right\}=0\label{phi}, \end{equation} where $g$ is the metric induced by
(\ref{eq:can}) on the hypersurfaces $t=const.$, $N=(-^4g^{tt})^{-\frac{1}{2}}$
is the lapse function and $\epsilon$ is the Levi-Civit\`a tensor on the
surfaces of constant time, $t$, and azimuthal angle, $\phi$. For simplicity the
whole equation has been written as the action of a differential operator, $L$,
on the potential.

This is a consequence only of the Maxwell equations in the curved spacetime
whose metric is given by Eq.~\ref{eq:can}, regardless of whether the
electromagnetic field is the source of the gravitational field.

Since we will be interested in compact sources, we shall only consider metrics
which are asymptotically flat in some coordinates $(t,r,\theta,\phi)$, 

\begin{eqnarray}ds^2&=&-\left(1-\frac{2m}{r}\right)\left(dt+\frac{2J\sin^{2}\theta}{r}d\phi\right)^2+
\nonumber\\&+&
\left(1+\frac{2m}{r}\right)\left\{dr^2+\left(r^2+c_1\,r\right)\left(d\theta^2
+\sin^2\theta d\phi^2\right)\right\}+O\left(\frac{1}{r^2}\right),
\end{eqnarray} where $m$ is the total mass of the source and $J$ is the total
angular momentum and $c_1$ is just a constant (its value is $-2m$ for
Kerr-Newman metrics). The electromagnetic potential of the compact source will
have the following asymptotic expansion,

\begin{equation}
\Phi=\frac{e}{r}+\frac{M\,\cos\theta}{r^2}+\frac{c_2}{r^2}+O(r^{-3}),
\end{equation} where $e$ is the total charge, $M$ is a complex constant whose
real and imaginary parts are, respectively, the electric and magnetic dipole
moment and $c_2$ is a constant that may arise in some choices of coordinates. 

As it was done in \cite{second}, \cite{first}, \cite{zip}, \cite{prep},  we shall introduce a function
$Z$ that satisfies the following elliptic differential equation out of the
surface source and behaves at infinity like the cartesian coordinate $z$,

\begin{equation} L^+\,Z=0\ \ \ \ \ \ \ \
Z=(r+c_3)\,\cos\theta+O(r^{-1})\label{zeda}, \end{equation} where $+$ denotes
the complex conjugate and $c_3$ is a constant.

The region where the electromagnetic surface source is located will be taken as
the surface, $S$, where the electromagnetic potential (or its derivative),
$\Phi$, is discontinuous, just as it is done in the classical theory of
potential. Without loss of generality we shall assume that this surface is
closed. The three-space, $\Omega$, defined by any of the hypersurfaces
$t=const.$, excluding $S$, will be divided into two regions, $\Omega^+$ and
$\Omega^-$, respectively the outer and inner parts of $\Omega$ with respect to
$S$. 

Taking into account Eq.~\ref{phi} and Eq.~\ref{zeda} the following Green
identity can be written, 

\begin{eqnarray}
0=\int_{\Omega}\sqrt{g}\,(Z\,L\Phi-\Phi\,L^+Z)\,dx^1dx^2dx^3=\nonumber\\=
\int_{\partial\Omega}\,dS\,N\left\{e^{-2U}\left(Z\,\frac{d\Phi}{dn}-\Phi\,\frac{dZ}{dn}
\right)+i\,\frac{A}{\rho}\,\left(Z\,*d\Phi(n)+\Phi\,*dZ(n)\right)\right\}, 
\end{eqnarray} using the Stokes theorem. The two-dimensional Hodge dual on the
surfaces of constant time and azimuthal angle is denoted by $*$ and $n$ is the
unitary outer normal to $S$.

The boundary $\partial\Omega^+$ is formed by $S$ and the sphere at infinity
whereas $\partial\Omega^-$ is the surface $S$. The integral at infinity can be
performed with the information we get from the asymptotic behaviours. The
discontinuity of the integrand on $S$,  \begin{equation}
\sigma_M=\frac{1}{4\pi}\,\left[N\left\{e^{-2U}\left(\Phi\,
\frac{dZ}{dn}-Z\,\frac{d\Phi}{dn}\right)
-i\,\frac{A}{\rho}\,\left(Z\,*d\Phi(n)+\Phi\,*dZ(n)\right)\right\}\right]
\label{magden}, \end{equation} denoted by squared brackets, can be interpreted
as the electromagnetic dipole moment surface density of the source for this
field. Its real part is the electric dipole density and its imaginary part is
the magnetic moment density.

 \section{The Kerr-Newman spacetime}

 As an example for this formalism, the magnetic dipole density for a surface in
the Kerr-Newman spacetime will be calculated. We shall follow \cite{is1} and
restrict the range of the Boyer-Lindquist coordinate $r$ to positive values.
Points on the hypersurface $r=0$ with coordinates $(t,\phi,0,\theta)$ and
$(t,\phi,0,\pi-\theta)$ are identified. Hence the Ernst potential, 

\begin{equation} \Phi=\frac{e}{r-i\,a\,\cos\theta}, \end{equation} will
be discontinuous on $r=0$.

 The Kerr-Newman metric in Boyer-Lindquist coordinates induces the line element,

\begin{equation}
ds^2_2=a^2\,\cos^2\theta\,d\theta^2+\sin^2\theta\,(a^2-e^2\,\tan^2\theta)\,d\phi^2,
\end{equation} on the surface $r=0$. We just need a solution of Eq.~\ref{zeda},
 
\begin{equation} Z=(r-2\,m)\,\cos\theta+\frac
{e^{2}\,\cos\theta+i\,a\,m\cos^2\theta}{r+i\,a\,\cos\theta}, \end{equation}  
in order to calculate the magnetic moment surface density,

\begin{equation} \sigma_M={\frac {\left
(e^{2}\,\cos^2\theta+e^{2}+a^{2}\,\cos^2\theta\right )
\,i\,e}{2\,\pi\,a^{2}\,\cos^3\theta\,\sqrt {a^{2}-e^{2}\,\tan^2\theta}}},
\end{equation}

\section*{Acknowledgments}
 The present work has been supported by DGICYT Project PB92-0183. The author
wishes to thank F. J. Chinea and L. M. Gonz\'alez-Romero for valuable
discussions.

 \end{document}